\documentclass[final,1p,times]{elsarticle}

\usepackage{color}
\usepackage{amsmath}
\usepackage{subfigure}
\usepackage{lineno,hyperref}

\journal{Physics of the Dark Universe}

\begin{document}
	
	\begin{frontmatter}
		
		\title{Comparing the scalar-field dark energy models with recent observations}

		\author[address1,address2]{Tengpeng Xu}
		
		\author[address1,address2]{Yun Chen\corref{mycorrespondingauthor}}
		\cortext[mycorrespondingauthor]{Corresponding author}
		\ead{chenyun@bao.ac.cn}
		
		\author[address3]{Lixin Xu}
		\author[address4]{Shuo Cao}
		
		\address[address1]{Key Laboratory for Computational Astrophysics, National Astronomical Observatories, Chinese Academy of Sciences, Beijing 100101, China}
		\address[address2]{College of Astronomy and Space Sciences, University of Chinese Academy of Sciences, Beijing, 100049, China}
		\address[address3]{Institute of Theoretical Physics, School of Physics, Dalian University of Technology, Dalian, 116024, China}
		\address[address4]{Gravitational Wave and Cosmology Laboratory, Department of Astronomy, Beijing Normal University, Beijing 100875, China}
		
		\begin{abstract}
		    In this paper, we investigate the general properties of a class of scalar-field dark energy models (i.e., $\phi$CDM models) which behave like cosmological trackers at early times. Particularly, we choose three $\phi$CDM models with typical potentials, i.e., $V(\phi)\propto \phi^{-\alpha}$ (inverse power-law (IPL) model), $V(\phi)\propto \coth^{\alpha}{\phi}$ (L-model) and $V(\phi)\propto \cosh(\alpha\phi)$ (Oscillatory tracker model), where the latter two models are based on the $\alpha$-attractors originated from the study of inflation. These models, which reduce to the $\Lambda$CDM model with $\alpha \to 0$, are studied and compared with the recent observations, including the Pantheon sample of type Ia supernovae (SNe Ia), baryon acoustic oscillations (BAO) measurements extracted from 6dFGS, BOSS and eBOSS, as well as the temperature and polarization anisotropy power spectra data of cosmic microwave background radiation (CMB) from $\emph{Planck}$ 2018 results. The observational constraints from the combining sample (SNe Ia + BAO + CMB) indicate that none of the three $\phi$CDM models exclude the $\Lambda$CDM model at $68.3\%$ confidence level. We find that the CMB anisotropy data have obvious advantages in constraining the dark energy models compared with other cosmological probes, which is particularly evident in the L-model. Furthermore, we apply the Bayesian evidence to compare the $\phi$CDM models and the $\Lambda$CDM model with the analysis of the combining sample. The concordance $\Lambda$CDM model is still the most supported one. In addition, among the three $\phi$CDM models, the IPL model is the most competitive one, while the L-model/Oscillatory tacker model is moderately/strongly disfavored. 
		\end{abstract}
		
		\begin{keyword}
			Dark energy; Quintessence; Cosmic microwave background; Baryon acoustic oscillations; Type Ia supernova
		\end{keyword}
		
	\end{frontmatter}
	
	
	\section{Introduction}
	The observations of type Ia supernovae (SNe Ia) showed that the universe is not only expanding but also accelerating \cite{astro-ph/9805201, 1201.2434, 1710.00845}. Subsequent observations, including the Cosmic Microwave Background Radiation (CMB) \cite{astro-ph/0302209, 1807.06209}, Baryon Acoustic Oscillations (BAO) \cite{0705.3323, 2007.08991}, gravitational lens statistics \cite{astro-ph/9912508, astro-ph/0209602, Cao2011, Cao2012, Cao2015, 1809.09845, Liu2019}, and milliarcsecond compact structure of radio quasars \cite{Cao2015b,Cao2017a,Cao2017b,Lian2021}, all support an accelerating expansion universe. Thus, the existence of dark energy becomes crucial, which is deemed as a cosmological component with negative pressure and can drive the cosmic expansion to accelerate. The $\Lambda$CDM model with the cosmological constant $\Lambda$ as a dark energy candidate is so far the simplest and most natural cosmological model, which can fit the above observations very well. Nevertheless, there are several problems in the $\Lambda$CDM model, such as the coincidence problem and the fine-tuning problem \cite{Weinberg:1989, 1001.1489}. The dilemma of the $\Lambda$CDM model mainly comes from the fact that the energy density of dark energy is constant in time. Naturally, we can anticipate the dark energy owning a time-varying energy density to solve or alleviate the problems. In practice, there are mainly two approaches to obtain the time-varying dark energy. One approach is to consider the dark energy with an evolving equation of state \cite{hep-th/0603057,Cao2014, 1910.09853,1205.3421,2006.00244,2007.08873,2105.07103}, and another is to consider the interaction between the dark energy and dark matter \cite{astro-ph/9908023, Cao2011b, Cao2013, 1908.04281}. 
	
	We focus on a popular kind of dynamical dark energy model, i.e., the scalar field dark energy models (or the quintessence models) \cite{Peebles:1988-1, Peebles:1988-2, astro-ph/9707286, astro-ph/9711102, astro-ph/9910097, astro-ph/9912046, astro-ph/9910214}, in which the dark energy is deemed to be a slowly evolving and spatially homogeneous scalar field. The cosmological model with the scalar-field dark energy and the cold dark matter is often called as $\phi$CDM model. The quintessence with some particular forms of potential $V(\phi)$ can own the ``tracker'' behavior \cite{astro-ph/9807002, astro-ph/9812313}, which enables the current energy density of dark energy to be reached from a wide range of initial conditions, and thus alleviates the coincidence problem. The quintessence with an inverse power-law (IPL) potential $V(\phi)\propto \phi^{-\alpha}$ \cite{Peebles:1988-1} is a canonical one with the ``tracker'' behavior, which has been widely studied with various data sets \cite{1312.1443,1309.3710,1805.06408,1902.03196,1105.5660,1106.4294,1507.02008,1603.07115}. However, the IPL model cannot make $w_{\phi}$ evolve below $-0.8$ at present stage, which is suggested by observations, and keep the ``tracker'' property at the same time \cite{1709.09193}. To solve this problem, Satadru et al. (2017) \cite{1709.09193} proposed a new class of quintessence models with ``tracker'' property, which are based on the $\alpha$-attractors family of potentials. The $\alpha$-attractors potential is originally introduced in the context of inflation by Kallosh and Linde \cite{1306.5220, 1311.0472}, then Swagat et al. (2017) \cite{1703.03295} and Satadru et al. (2017) \cite{1709.09193} extend the study and find that the $\alpha$-attractors can give rise to new dark energy models with ``tracker'' properties. Four different quintessence models are described in \cite{1709.09193}, whereas only two prominent examples among them are taken into account in this work, including the scenarios of  $V(\phi)\propto \coth^{\alpha}{\phi}$ (L-model) and $V(\phi)\propto \cosh(\alpha\phi)$ (Oscillatory tracker model). 
	
	In this work, we focus on three $\phi$CDM models with different ``tracker'' potentials, i.e., IPL model, L-model and Oscillatory tracker model, by using three types of the recent observational data: the Pantheon sample \cite{1710.00845} of SNe Ia ($0.01 < z < 2.26$) (SNe Ia data), the recent BAO sample ($0.106 < z < 2.36$) collected in \cite{1805.06408} (BAO data), and the CMB temperature and polarization anisotropies data from the \emph{Planck}-2018 survey (\emph{Planck} data) \cite{1807.06209}. The rest of the paper is organized as follows. In section 2, we briefly introduce the scalar field dark energy and the $\phi$CDM models under consideration. In section 3, we describe the data samples used in this work, i.e., the SNe Ia data, BAO data and \emph{Planck} data. In section 4, we put observational constraints on the model parameters of the three $\phi$CDM models, and apply the Bayesian evidence to compare the $\phi$CDM models with the $\Lambda$CDM model. In the last section, we summarize the main conclusions.
	
	\section{$\phi$CDM models with tracker potentials}
	
	In the framework of Einstein's theory of general relativity, the universe can be described by the action:
	\begin{equation}
		S=\int{\sqrt{-g}(-\frac{m_\mathrm{p}^2}{16\pi}R+\mathcal{L}_{\mathrm{m,r}}+\mathcal{L}_{\mathrm{DE}})\mathrm{d}^4x},
		\label{action}
	\end{equation}
	where $g$ is the determinant of the metric $g_{\mu\nu}$, $R$ is the Ricci scalar, $m_\mathrm{p}=1/\sqrt{G}$ is the Planck mass with $G$ being the Newtonian constant of gravitation, $\mathcal{L}_{m,r}$ is the Lagrangian density of matter and radiation, while $\mathcal{L}_{\mathrm{DE}}$ is the Lagrangian density of dark energy.
	In the quintessence scalar field model of dark energy, the Lagrangian density of dark energy can be written as:
	\begin{equation}
		\mathcal{L}_{\mathrm{\phi}}=\frac{1}{2}g^{\mu\nu}\partial_{\mu}\phi\partial_{\nu}\phi-V(\phi),
		\label{Lagrangian phi}
	\end{equation}
	where $V(\phi)$ is the potential of the field $\phi$. Here we first specify the dimensions of these physical quantities. Planck units are adopted in this work, i.e., $c=\hbar=1$, hence the dimension of $V(\phi)$ is $[V(\phi)]=[m_{\mathrm{p}}^2]=[t^{-2}]$, while $\phi$ is dimensionless (in some other papers, $\phi$ field has the same unit with $m_{\rm p}$, here we make it dimensionless). For convenience, the current Hubble parameter $H_0=1$ is also used in the following derivation.
	
	The Eq.(\ref{Lagrangian phi}) implies that there are two variables in the action, the metric $g_{\mu\nu}$ and the scalar field $\phi$. According to the least action principle, the Einstein equation and the motion equation of the $\phi$ field can be derived by variations with respect to $g_{\mu\nu}$ and $\phi$, respectively. One can work out the energy-momentum tensor of the $\phi$ field
	\begin{eqnarray}
		T_{\mu\nu}&=&-2\frac{\delta\mathcal{L}_{\phi}}{\delta g^{\mu\nu}}+\mathcal{L}_{\phi}g_{\mu\nu} \nonumber \\
		&=& \partial_{\mu}\phi\partial_{\nu}\phi-g_{\mu\nu}\left[\frac{1}{2}g^{\alpha\beta}\partial_{\alpha}\phi\partial_{\beta}\phi+V(\phi)\right].
	\end{eqnarray}
	Then the energy density and pressure of the field $\phi$ can be written as:
	\begin{eqnarray}
		\rho_{\phi}&=&\frac{1}{2}\dot{\phi}^2+V(\phi)
		\label{rho_phi}
		\\
		p_{\phi}&=&\frac{1}{2}\dot{\phi}^2-V(\phi),
	\end{eqnarray}
	and the equation of state (EoS) of field $\phi$:
	\begin{equation}
		w_{\phi}=\frac{p_{\phi}}{\rho_{\phi}}=\frac{\dot{\phi}^2-2V(\phi)}{\dot{\phi}^2+2V(\phi)}.
	\end{equation}
	
	In the universe with a Friedmann-Lemaitre-Robertson-Walker(FLRW) metric, by substituting $g_{\mu\nu}$, $T_{\mu\nu}$ and $R_{\mu\nu}$ into the Einstein equation, one can obtain the equation of motion of field $\phi$, i.e., the Klein-Gordon equation,
	\begin{equation}
		\ddot{\phi}+3(\frac{\dot{a}}{a})\dot{\phi}+\frac{\mathrm{d}V}{\mathrm{d}\phi}=0,
		\label{KG_eq}
	\end{equation}
	and the Friedmann equation
	\begin{equation}
		H^2(z)=\frac{8\pi}{3m_{\mathrm{p}}^2}(\rho_{\mathrm{m}}+\rho_{\phi})-\frac{K}{a^2},
		\label{FM_eq}
	\end{equation}
	where $a$ is the scale factor, $H \equiv \dot{a}/a$ is the Hubble parameter, $z$ is the cosmological redshift, $\rho_{\mathrm{m}}$ is the matter density and $K$ is the spatial curvature, while $K>0,K<0$ and $K=0$ correspond to closed, open and flat universe, respectively. Considering the early universe dominated by matter, one can derive the matter density as \cite{1309.3710}
	\begin{equation}
		\rho_{\mathrm{m}}=\rho_{\mathrm{m0}}a^{-3}.
		\label{rho_m}
	\end{equation}
	Substituting Eq.(\ref{rho_phi}) and Eq.(\ref{rho_m}) into Eq.(\ref{FM_eq}) and combining it with Eq.(\ref{KG_eq}), one can obtain the differential equations of $\phi$ and $a$ with respect to $t$.
	
	Once the potential function $V(\phi)$ is determined, the differential equations above can be solved, and the Hubble parameter $H(z)$ can be computed numerically. However, it’s impractical to determine a specific functional form of $V (\phi)$ based solely on the fundamental physics. Hence a number of phenomenological forms for $V(\phi)$ have been proposed over the past three decades.
	From the theoretical perspective, the distinguished forms for $V(\phi)$ should satisfy the following two conditions: (i) it must be able to drive the late-time acceleration of the universe ; (ii) it should possess the ``tracker'' property that enables the current density of the dark energy to be reached from a wide range of initial conditions, such that the coincidence problem can be alleviated.
	Three parametric forms of $V(\phi)$, which satisfy the two conditions mentioned above, are considered in this work.
	The first model is the inverse-power law  (IPL) model,
	\begin{equation}
		V(\phi)=\hat{V} \phi^{-\alpha},
	\end{equation}
	which is the simplest and the best-studied $\phi$CDM model.
	The second model is the L-model \cite{1709.09193},
	\begin{equation}
		V(\phi)=\hat{V} \mathrm{coth}^{\alpha}(\phi).
	\end{equation}
	It is called as the L-model since $V(\phi)$ has a characteristic ``L'' shape. The L-model introduced in Satadru et al. (2017) \cite{1709.09193} is expressed as $V(\phi)=\hat{V}\mathrm{coth}^p(\lambda\phi)$. In order to maintain the same degree of freedom as other $\phi$CDM models, we set $\lambda=1$ and rename the free parameter $p$ as $\alpha$ in this work.
	At early times of the universe, $\phi<<1$, one has $V(\phi)\simeq \hat{V} \phi^{-\alpha}$, which implies that the behavior of this model is very similar to that of the IPL model at early times. 
	At late times of the universe, $\phi>>1$, one reads
	\begin{equation}
		V(\phi)=\hat{V} \mathrm{coth}^{\alpha}(\phi)\simeq \hat{V}[1+2e^{-2\phi}]^{\alpha}\simeq \hat{V}[1+2\alpha e^{-2\phi}].
		\label{eq:Vphi_V0}
	\end{equation}
	Thus, when $t\rightarrow \infty$, one can get $V(\phi)\rightarrow \hat{V}$ from Eq.(\ref{eq:Vphi_V0}), so the Lagrangian $\mathcal{L}_{\phi}$ converges to a nonzero constant in the far future, that means the L-model behaves like a cosmological constant at late times. Especially, while $\alpha=2$, the L-model has a special importance in the nDGP braneworld scenario, which can cause $\Lambda$CDM-like background expansion on the phantom brane \cite{1807.00684}.
	The third model under consideration is the Oscillatory tracker model \cite{1709.09193},
	\begin{equation}
		V(\phi)=\hat{V} \mathrm{cosh}(\alpha \phi).
	\end{equation}
	At early times, $|\phi| >>1$, its asymptotic form is:
	\begin{equation}
		V(\phi)\sim\frac{\hat{V}}{2}\mathrm{exp}(|\alpha \phi|),
	\end{equation}
	which implies that the behavior of this model is similar to that of the exponential potential in the early epoch.
	At late times, $0<|\phi|<<1$, its asymptotic form becomes:
	\begin{equation}
		V(\phi)\sim \hat{V}[1+\frac{1}{2}(\alpha\phi)^2],
	\end{equation}
	which indicates that $V(\phi)$ will gradually approach the constant $\hat{V}$, and $w_{\phi}$ will approach -1. In addition, because of the existence of $\phi^2$, $w_{\phi}$ will be oscillatory around -1 in the late epoch; hence this model is named as Oscillatory tracker model.
	
	In the frameworks of these three $\phi$CDM models with tracker properties, we can solve the differential equations above by setting the initial condition $\phi_{\mathrm {i}}$ and $\dot{\phi}_{\mathrm {i}}$ at comparatively arbitrary $t_{\mathrm {i}}$ and work out the expansion history of the universe. To achieve this, we need four parameters  $\textbf{p}\equiv(\Omega_{\mathrm{m}0}, \Omega_{\mathrm{k}0}, \hat{V},\alpha)$. The first two are the cosmological geometry parameters, where 0 denotes the present epoch, while the last two parameters are the model parameters in $V(\phi)$. According to the dimensional analysis $[\hat{V}]=[m_{\rm p}^2]=[t^{-2}]=[H_0^2]$, we have $\hat{V} \propto H_0^2$. Because of $H_0=1$ as mentioned previously, $\hat{V}$ can be reduced during the calculation, then we only need three parameters $\textbf{p}\equiv(\Omega_{\mathrm{m}0}, \Omega_{\mathrm{k}0}, \alpha)$.
	Then the comoving distance $d_\mathrm{C}(z)$ can be calculated by:
	\begin{equation}
		\label{dC}
		d_{\mathrm C}(\textbf{p};z)=\frac{c}{H_0}\int^z _0\frac{\mathrm{d}z^\prime}{E(\textbf{p};z^\prime)}=\frac{c}{H_0}\int_{t_z}^{t_0}\frac{\mathrm{d}t}{a(\textbf{p};t)},
	\end{equation}
	where $E(z)=H(z)/H_0$, $t_z$ is the cosmic time at redshift $z$. Meanwhile, the  transverse comoving distance $d_{\mathrm M}$ can be worked out with the following expression:
	\begin{equation}
		\label{dM}
		d_{\mathrm M}(\textbf{p};z) = \left\{
		\begin{aligned}
			&  d_{\mathrm C}(\textbf{p};z) & & \text{if}\ \Omega_{k0} = 0, \\
			&  \frac{c}{H_0\sqrt{\Omega_{k0}}}\mathrm{sinh}[\sqrt{\Omega_{k0}}H_0d_{\mathrm C}(\textbf{p};z)/c] & & \text{if}\ \Omega_{k0} > 0, \\
			&  \frac{c}{H_0\sqrt{\left|\Omega_{k0}\right|}}\mathrm{sin}[\sqrt{\left|\Omega_{k0}\right|}H_0d_{\mathrm C}(\textbf{p};z)/c] & & \text{if}\ \Omega_{k0} < 0.
		\end{aligned}
		\right.
	\end{equation}
	
	\section{Observational data sets}
	
	\subsection{The Supernovae data}
	The recent Pantheon sample of SNe Ia including 1048 data points with $0.01 < z < 2.26$ \cite{1710.00845}, which has been applied to the constraints of numerous cosmological models \cite{1801.02371}, is also employed in our cosmological analysis.
	The corrected apparent magnitude is used as an observable quantity for each Pantheon supernova (see Table A17 of \citep{1710.00845}), i.e.,
	\begin{equation}
		\begin{aligned}
			m_{\rm obs}&\equiv\mu+M \\
			&=m_{\mathrm{B}}+(\alpha x_1-\beta c+\Delta_{\mathrm{M}}+\Delta_{\mathrm{B}}),
		\end{aligned}
	\end{equation}
	where $\mu$ and $m_B$ are the distance modulus and the apparent magnitude of B-band, respectively. $x_1$ is the light-curve shape parameter and $c$ is the color parameter, then $M$ is the absolute magnitude of B-band in the fiducial SNe Ia model based on $x_1=0,c=0$. $\alpha$ is the coefficient of the relation between luminosity and stretch, $\beta$ is the coefficient of the relation between luminosity and color. What's more, $\Delta_{\mathrm{M}}$ is a distance correction based on the host-galaxy mass of the supernova, and $\Delta_{\mathrm{B}}$ is a distance correction based on predicted biases from
	simulations. 
	
	The theoretical prediction of distance modulus $\mu$ is expressed as:
	\begin{eqnarray}
		\mu_{\mathrm{th}}(\textbf{p};z)&=&5\log_{10}\left[\frac{d_{\mathrm L}(\textbf{p};z)}{\mathrm{Mpc}}\right]+25\nonumber \\
		&=&5\log_{10}\left[D_L(\textbf{p};z)\right]+\mu_0,
	\end{eqnarray}
	where $d_{\mathrm{L}}(\textbf{p};z)=(1+z)d_{\mathrm{M}}(\textbf{p};z)$ is the luminosity distance, 
	$D_{\mathrm L}\equiv d_{\mathrm L}/(\frac{c}{H_0})$ is the dimensionless luminosity distance, and $\mu_0 = 5\log_{10}(\frac{cH_0^{-1}}{\mathrm{Mpc}})+25$. 
	Thus the corresponding theoretical apparent magnitude is:
	\begin{equation}
		m_{\mathrm{th}}(\textbf{p};z)=\mu_{\mathrm{th}}(\textbf{p};z)+M=5\log_{10}[D_{\mathrm L}(\textbf{p};z)]+M^\prime,
		\label{eq:m_th}
	\end{equation}
	where $M^\prime=\mu_0+M$ is a term related to the Hubble constant $H_0$ and the absolute magnitude $M$.
	
	The likelihood for the Pantheon sample is $\mathcal{L}_{\rm SNe}\propto \exp(-\frac{\chi_{\rm SNe}^2}{2})$, where
	\begin{equation}
		\chi^2_{\rm {SNe}}(\textbf{p})=\sum_{i,j}\Delta m_i(\textbf{p}) C^{-1}_{ij} \Delta m_j(\textbf{p}),
		\label{eq:chi2_SNe}
	\end{equation}
	where $\Delta m_i(\textbf{p})=m_{\mathrm{th}}(\textbf{p};z_i)-m_{\mathrm{obs},i}$, and $C_{ij}$ is the covariance matrix which includes the contributions from both the statistical and systematic errors. Following the method presented in Giostri et al.(2012) \cite{1203.3213}, we conduct the analytic marginalization over the term $M^\prime$ which is composed of $H_0$ and $M$, since they are both uninterested in this work.
	
	\subsection{The BAO data}
	
	We use a sample of 11 BAO data within $0.106 < z < 2.36$ collected in Ryan et al.(2019) \cite{1805.06408}, which contains six correlated measurements from SDSS DR12 \cite{1607.03155} and five uncorrelated measurements from 6dFGS \cite{1106.3366}, SDSS DR7 \cite{1409.3242}, SDSS DR11 \cite{1311.1767}, SDSS DR12 \cite{1702.00176} and SDSS DR14 \cite{1705.06373}. The six correlated measurements are the scaled transverse comoving distance $d_{\mathrm M}(r_{\mathrm {s,fid}}/r_{\mathrm s})$ and the scaled Hubble parameter $H(z)(r_{\mathrm s}/r_{\mathrm {s,fid}})$ at redshift $z = 0.38$, 0.51 and 0.61, respectively. The five uncorrelated measurements are $r_{\mathrm s}/d_{\mathrm V}(z=0.106)$, $d_{\mathrm V}(r_{\mathrm s}/r_{\mathrm {s,fid}})(z=0.15)$, $d_{\mathrm V}(r_{\mathrm s}/r_{\mathrm {s,fid}})(z=1.52)$, $\frac{d_{H}^{0.7}d_{M}^{0.3}}{r_s}(z=2.33)$, and $c/(r_s H(z=2.36))$, where
	$d_{\mathrm V}$ is the volume-averaged angular diameter distance
	\begin{equation}
		d_{\mathrm V}(\textbf{p}; z) = \left[\frac{cz}{H_0}\frac{d_{\mathrm M}^2(\textbf{p}; z)}{E(\textbf{p};z)}\right]^{1/3},
	\end{equation}
	$r_s$ and $r_{\mathrm {s,fid}}$ are the size of the sound horizon at the drag epoch and the corresponding value calculated with the fiducial cosmology, respectively.
	
	All these BAO measurements are scaled by the sound horizon $r_s$ at the baryon drag epoch $z_d$, which is expressed as:
	\begin{equation}
		\begin{aligned}
			r_s &= \int^{\eta(z_d)}_0 c_s(1+z) \rm d \eta, \\
			c_s &= 1/\sqrt{3(1+R)}, \\
			R &= 3\rho_b/4\rho_\gamma,
		\end{aligned}
	\end{equation}
	where $c_s$ is the sound speed and $\eta(z)$ is the conformal time at the redshift $z$. To find the drag epoch redshift $z_d$, one can numerically solve the following integral equation \cite{1003.3999, 1208.3715}
	\begin{equation}
		\begin{aligned}
			\tau(\eta_d)&\equiv\int^{\eta_0}_{\eta}\dot\tau_d\rm d \eta^\prime\\
			&=\int^{z_d}_0\frac{x_e(z)\sigma_T}{R}\frac{\rm d\eta}{\rm d a}\rm dz = 1
		\end{aligned}
	\end{equation}
	where $\sigma_T$ is the Thomson cross-section and $x_e(z)$ is the fraction of free electrons. These numerical calculation are embedded in the code   \href{https://camb.info/}{\textbf{CAMB}}\footnote{https://camb.info/} \cite{astro-ph/9911177} which is applied in our analysis.
	
	The likelihood for the BAO sample is $\mathcal{L}_{\rm BAO}\propto\exp(-\frac{\chi^2_{\rm BAO}}{2})$, where $\chi^2_{\rm BAO}$ is constructed as
	\begin{equation}
		\begin{aligned}
			\chi^2_{\rm BAO}(\textbf{p})=&\sum_{i,j}[\mathcal{A}_{\rm th}(\textbf{p};z_i)-\mathcal{A}_{\rm obs}(z_i)] C^{-1}_{ij} [\mathcal{A}_{\rm th}(\textbf{p};z_j)-\mathcal{A}_{\rm obs}(z_j)] + \\ &\sum_k\frac{[\mathcal{A}_{\rm th}(\textbf{p};z_k)-\mathcal{A}_{\rm obs}(z_k)]^2}{\sigma_k^2},
			i,j=1,2,...,6;k=1,2,...,5
		\end{aligned}
		\label{eq:chi2_BAO}
	\end{equation}
	where the two terms correspond to the correlated and uncorrelated measurements, respectively, $\mathcal{A}$ represents each BAO measurement and $C_{ij}^{-1}$ in the first term is the inverse of the covariance matrix, which can be found from \cite{1805.06408}. 
	
	\subsection{The \emph{Planck} CMB data}
	
	We employ the angular power spectra of temperature anisotropies and polarization of the CMB, including the TT, EE and TE power spectra data, to constrain the cosmological models. Correspondingly, where the code \href{https://camb.info/}{\textbf{CAMB}} \cite{astro-ph/9911177} is used to calculate the theoretical CMB spectra for each cosmological model. 
	
	The \emph{Planck} 2018 baseline likelihood release consists of a code package\footnote{\textit{COM\_Likelihood\_Code-v3.0\_R3.10.tar.gz}} and a baseline data package\footnote{\textit{COM\_Likelihood\_Data-baseline\_R3.00.tar.gz}} which can be downloaded from \href{http://pla.esac.esa.int/pla/}{Planck Legacy Archive}\footnote{http://pla.esac.esa.int/pla/}. The code complies to a library, allowing for the computation of log likelihoods for a given data set. The baseline data package contains ten data sets, six of which belong to the high-$l$ part ($l\geq 30$) and four belong to low-$l$ part ($2\leq l \leq 29$). We introduce the data sets used in this work as follows.
	\begin{itemize}
		\item low-$l$ TT - \textbf{Commander}
		
		This data set allows for the computation of the CMB TT likelihood in the range $l=2-29$. The likelihood is based on the results of the Commander approach \cite{1907.12875}, which implements a Bayesian component-separation method in pixel space, sampling the posterior distribution of the parameters of a model that describes both the CMB and the foreground emissions in a combination of the Planck maps. The samples of this exploration are used to infer the foreground-marginalized low-$l$ likelihood for any TT CMB spectrum.
		\item low-$l$ EE - \textbf{SimAll}
		
		This data set allows for the computation of the EE likelihood in the range $l=2-29$. The likelihood is estimated by comparing a cross-quasi-maximum-likelihood algorithm (QML) on the 100- and 143-GHz maps to high fidelity end-to-end simulations of the HFI instrument, as described in \cite{1907.12875}.
		\item high-$l$ TT+TE+EE - \textbf{Plik}
		
		This data set allows for the computation of the CMB joint TT, TE, and EE likelihood in the range $l=30-2508$ for TT and $l=30-1996$ for TE and EE. The file contains the 100-GHz, 143-GHz, and 217-GHz binned half-mission T and E cross-spectra. In temperature, only the $100\times100$, $143\times143$, $143\times217$, and $217\times217$ spectra are actually used, while in TE and EE all of them are used. Masks and multipole ranges for each spectrum are different and described in \cite{1907.12875}.
	\end{itemize}
	
	Following the $\emph{Planck}$ 2018 paper \cite{1807.06209}, we utilize the \emph{Planck} TTTEEE+lowE likelihood in our analysis, which is the combination of the baseline \textbf{Plik} likelihood, the \textbf{Commander} likelihood and the \textbf{SimAll} likelihood. Hereafter the likelihood for the \emph{Planck} data $\mathcal{L}_{\emph{Planck}}$ denotes the \emph{Planck} TTTEEE+lowE likelihood.
	
	\section{Observational constraints and model comparison}
	
	\begin{figure}[h!]
		\centering
		\large IPL model\\
		\subfigure{\includegraphics[width=\textwidth]{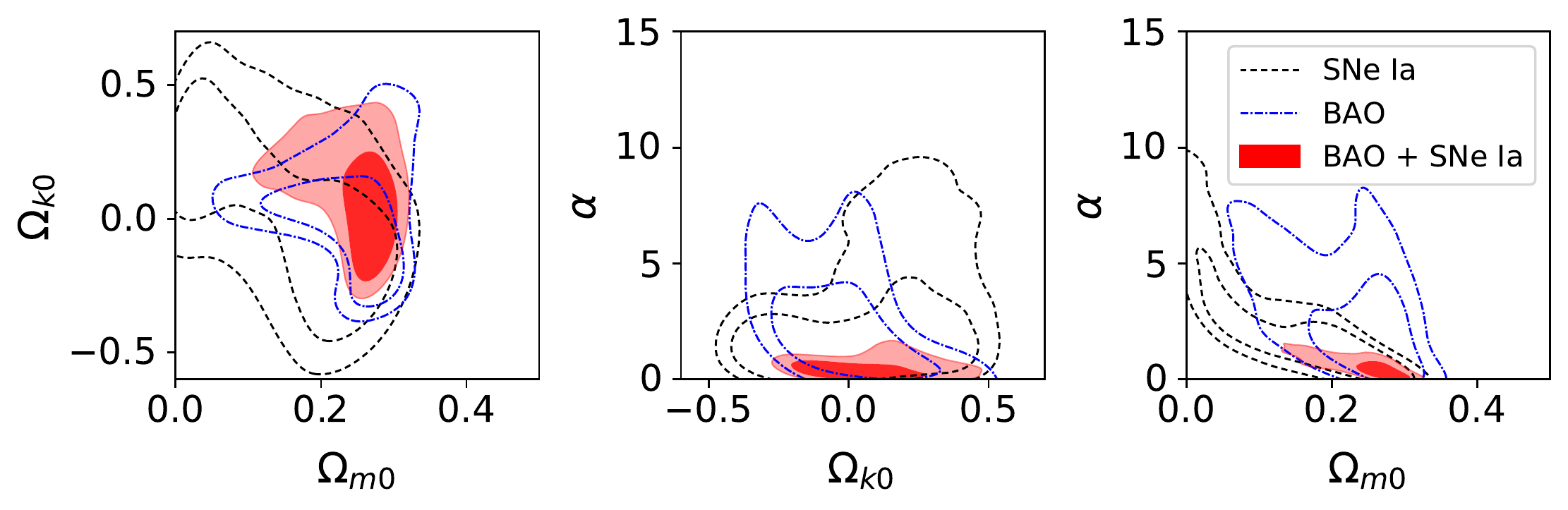}}\\
		L-model\\
		\subfigure{\includegraphics[width=\textwidth]{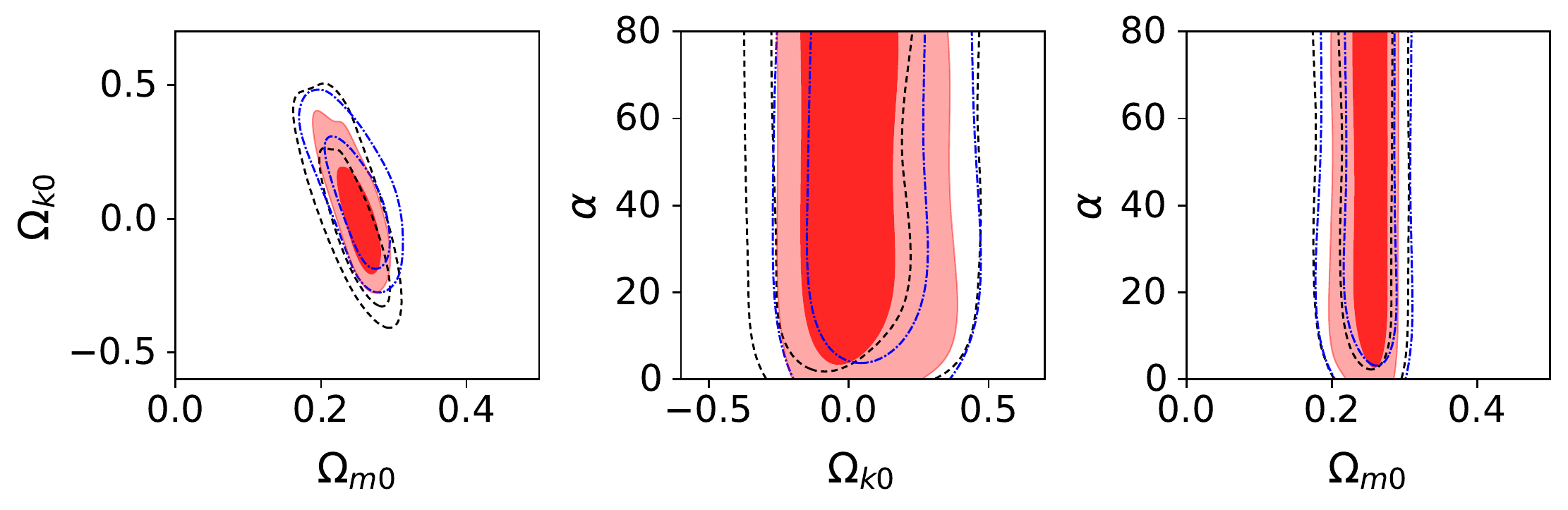}}\\
		Oscillatory tracker model \\
		\subfigure{\includegraphics[width=\textwidth]{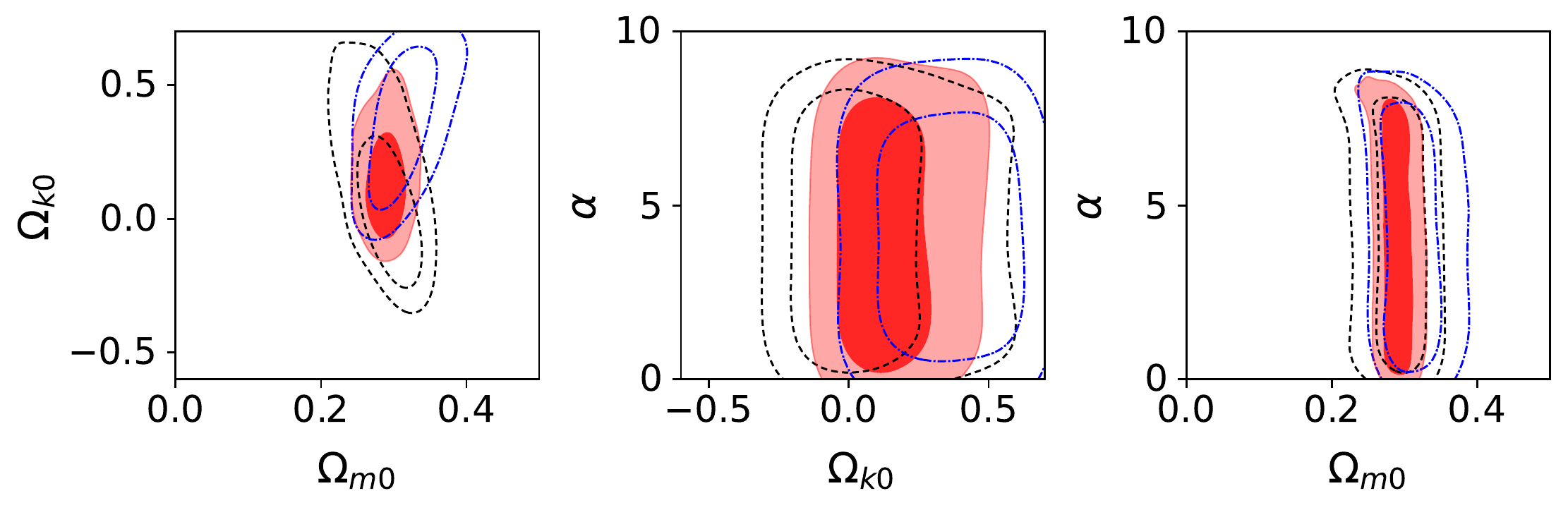}}\\
		
		\caption{Confidence intervals at $68.3\%$ and $95.4\%$ in the  ($\Omega_{\rm m0}, \Omega_{\mathrm{k0}}$), ($\Omega_{\rm k0}, \alpha$),  and ($\Omega_{\mathrm{m0}}, \alpha$) planes for the IPL model, L-model and Oscillatory tracker model, respectively, arising from the SNe Ia and BAO data.}
		
		\label{fig:results} 
	\end{figure}
	
	\begin{figure}[h!]
		\centering
		\large IPL model\\
		\subfigure{\includegraphics[width=\textwidth]{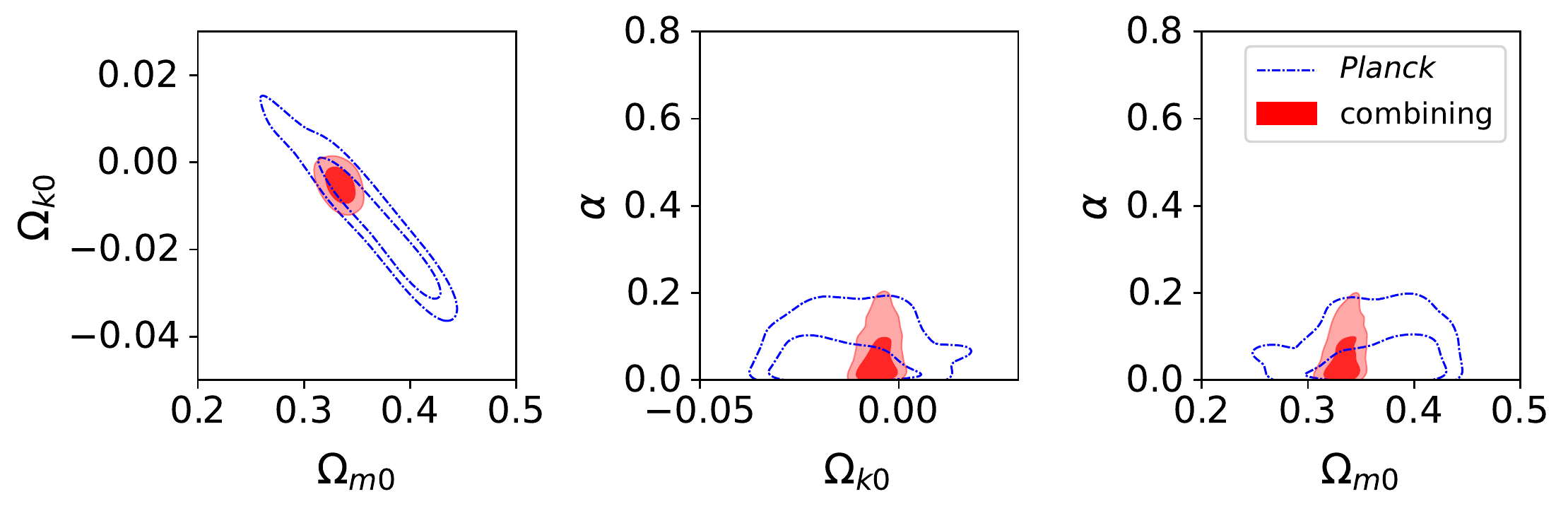}}\\
		L-model\\
		\subfigure{\includegraphics[width=\textwidth]{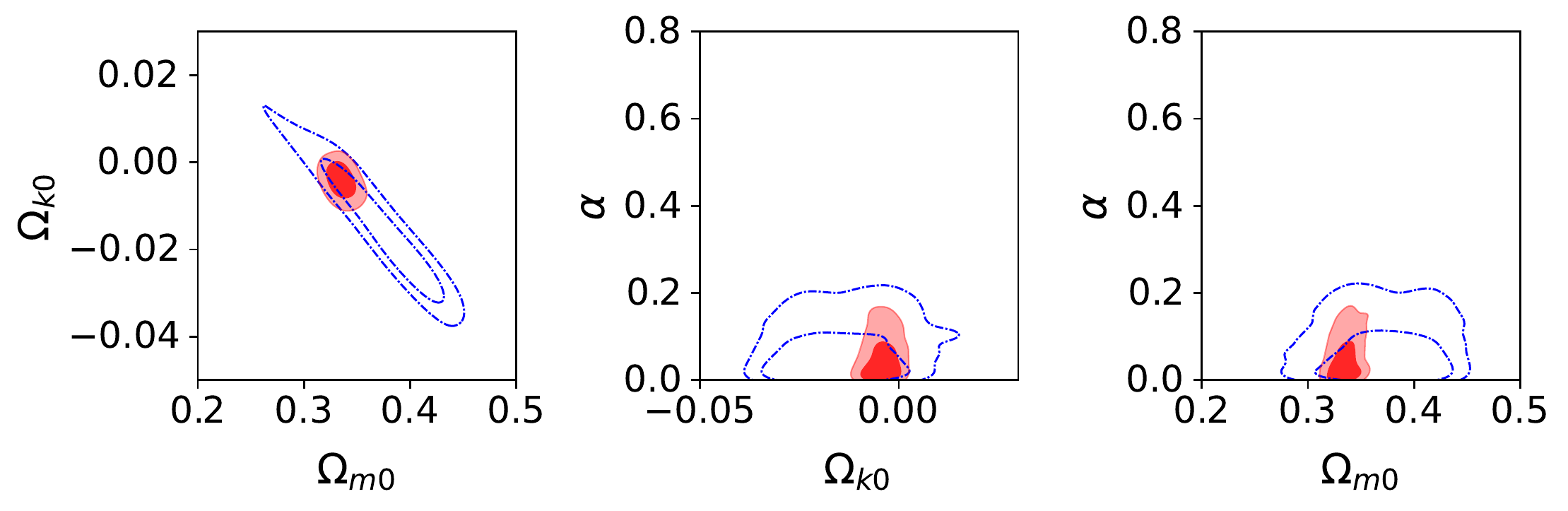}}\\
		Oscillatory tracker model \\
		\subfigure{\includegraphics[width=\textwidth]{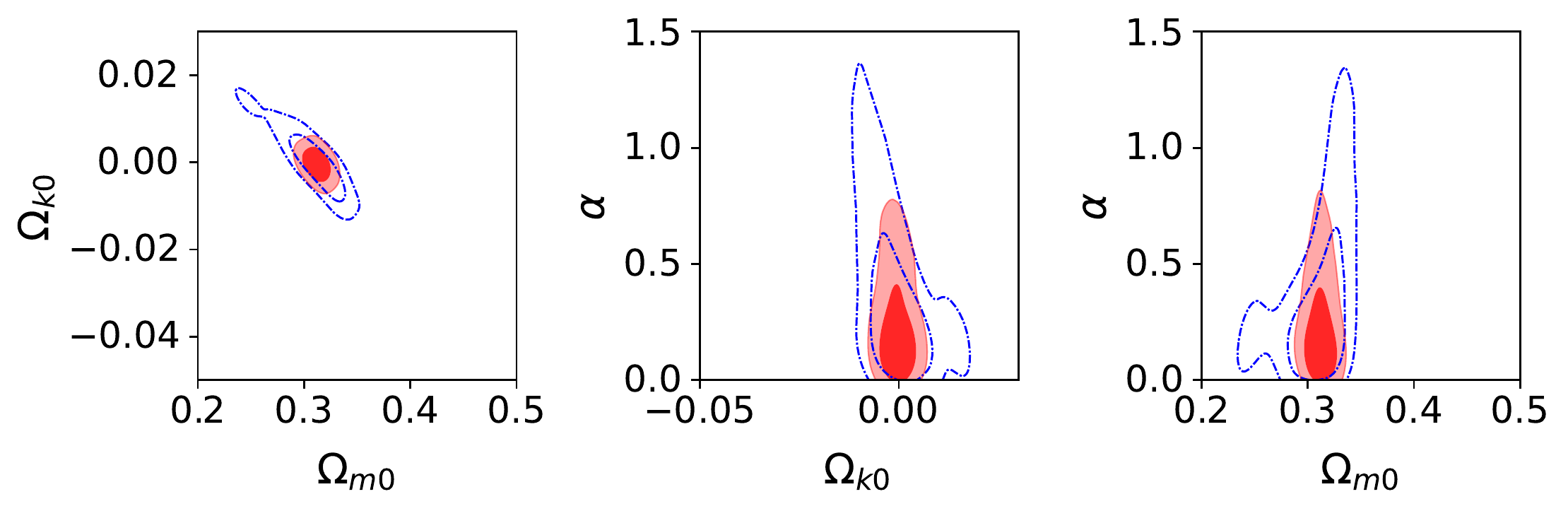}}\\
		
		\caption{Confidence intervals at $68.3\%$ and $95.4\%$ in the  ($\Omega_{\rm m0}, \Omega_{\mathrm{k0}}$), ($\Omega_{\rm k0}, \alpha$),  and ($\Omega_{\mathrm{m0}}, \alpha$) planes for the IPL model, L-model and Oscillatory tracker model, respectively, arising from the \emph{Planck} data and the combining sample (SNe Ia+BAO+\emph{Planck}).}
		
		\label{fig:results_planck} 
	\end{figure}
	
	\begin{table*}[h!]
		\small
		\begin{center}
			\begin{tabular}{|c|c|c|c|c|c|c|c|}
				\hline  
				model & data & $\Omega_{\rm m0}$ & $\Omega_{\mathrm{k} 0}$ & $\alpha$ & $\chi^2_{\rm{eff, min}}$  & $\ln{B_i}$ & $\ln{B_{i0}}$\\ 
				\hline  
				&&&&&&&\\ 
				& SNe Ia & $0.380^{+0.170}_{-0.150}$ & $-0.110^{+0.190}_{-0.240}$ & -&- &-  &-\\
				&&&&&&&\\
				$\mathcal{M}_0$ & BAO & $0.246^{+0.035}_{-0.039}$ & $0.200\pm0.150$ & -&- &- &- \\ 
				&&&&&&& \\
				& SNe Ia+BAO & $0.264\pm0.030$ & $0.102^{+0.079}_{-0.097}$ & - & 1034.32 & - & - \\
				&&&&&&& \\
				& $\emph{Planck}$ & $0.327^{+0.018}_{-0.014}$& $-0.026^{+0.012}_{-0.038}$& -& 2768.84 &-& - \\ 
				&&&&&&& \\
				& combining & $0.3258^{+0.0087}_{-0.0097}$& $-0.0033^{+0.0013}_{-0.0027}$& -& 3865.49& -1923.57&0 \\ \hline 
				&&&&&&&\\
				& SNe Ia & $<0.216$ & $0.030^{+0.290}_{-0.330}$ & $<2.527$  &-&-&- \\
				&&&&&&& \\
				$\mathcal{M}_1$ & BAO & $0.241^{+0.066}_{-0.023}$ & $-0.010^{+0.130}_{-0.200}$ & $<2.766$ &-&-&- \\ 
				&&&&&&&\\
				& SNe Ia+BAO & $0.256^{+0.038}_{-0.011}$ & $0.050^{+0.130}_{-0.180}$ & $<0.538$ & 1033.83 & - & - \\
				&&&&&&& \\
				& $\emph{Planck}$ & $0.369^{+0.047}_{-0.033}$& $-0.0143^{+0.0093}_{-0.0130}$& $<0.064$ &2767.52&-&- \\ 
				&&&&&&&\\
				& combining & $0.3340\pm{0.0094}$& $-0.0052\pm{0.0028}$& $<0.060$&3863.09 & -1924.23  &-0.66\\ 
				\hline 
				&&&&&&&\\
				& SNe Ia & $0.245^{+0.035}_{-0.025}$ & $0.000^{+0.140}_{-0.260}$ & $>0$&- & - &- \\ 
				&&&&&&&\\
				$\mathcal{M}_2$ & BAO & $0.251^{+0.033}_{-0.023}$ & $0.070^{+0.130}_{-0.190}$ & $>0$ & -&-&- \\ 
				&&&&&&&\\
				& SNe Ia+BAO & $0.250^{+0.024}_{-0.015}$ & $0.016^{+0.098}_{-0.170}$ & $>0$ & 1033.46 & - & - \\
				&&&&&&&\\
				& $\emph{Planck}$ & $0.374^{+0.045}_{-0.032}$& $-0.0155^{+0.0095}_{-0.0130}$& $<0.075$ &2767.85&-&- \\ 
				&&&&&&&\\
				& combining & $0.3350\pm{0.0091}$& $-0.0042\pm{0.0028}$& $<0.054$&3873.34  &-1928.57&-5.00   \\ 
				\hline
				&&&&&&& \\
				& SNe Ia & $0.290^{+0.035}_{-0.023}$ &  $0.070^{+0.120}_{-0.250}$ & $<5.720$ & -&-&- \\ 
				&&&&&&&\\
				$\mathcal{M}_3$ & BAO & $0.308\pm{0.044}$ & $0.350^{+0.220}_{-0.180}$ & $<5.600$  &-&-&- \\
				&&&&&&&\\
				& SNe Ia+BAO & $0.282^{+0.025}_{-0.010}$ & $0.140^{+0.100}_{-0.160}$ & $<5.914$ & 1033.93 & - & - \\
				&&&&&&&\\
				 & $\emph{Planck}$ & $0.311^{+0.022}_{-0.013}$& $-0.0003^{+0.0045}_{-0.0057}$& $<0.338$ &2768.27&- &-  \\
				&&&&&&&\\
				& combining & $0.3119\pm{0.0086}$& $-0.0004\pm{0.0026}$& $<0.233$&3865.77 & -1925.46&-1.89   \\ 
				\hline
			\end{tabular}
		\end{center}
		\caption{ Constraints on the $\Lambda$CDM model ($\mathcal{M}_0$) and the three $\phi$CDM models ($\mathcal{M}_1$, $\mathcal{M}_2$ and $\mathcal{M}_3$ denote IPL model, L-model and Oscillatory tracker model, respectively) from the SNe Ia, BAO and $\emph{Planck}$ data. The marginalized means and $68.3\%$ confidence interval of $\Omega_{\mathrm m 0}$ and $\Omega_{\mathrm k 0}$ obtained from MCMC analysis are given in the first two numerical columns, while the $68.3\%$ confidence interval for $\alpha$ which is positive definite in the three models are shown in the third numerical column. Correspondingly, we list the minimum effective $\chi^2$, the natural logarithm of the Bayesian evidences $\ln{B_i}$ and the Bayes factors $\ln{B_{i0}}$ of the combining sample in the last three columns, where $i=1,2,3$ and 0 denote the three $\phi$CDM models the $\Lambda$CDM model, respectively. }
		\label{tab:results}
	\end{table*}
	
	\begin{figure}[h!]
		\centering
		\includegraphics[width=\textwidth]{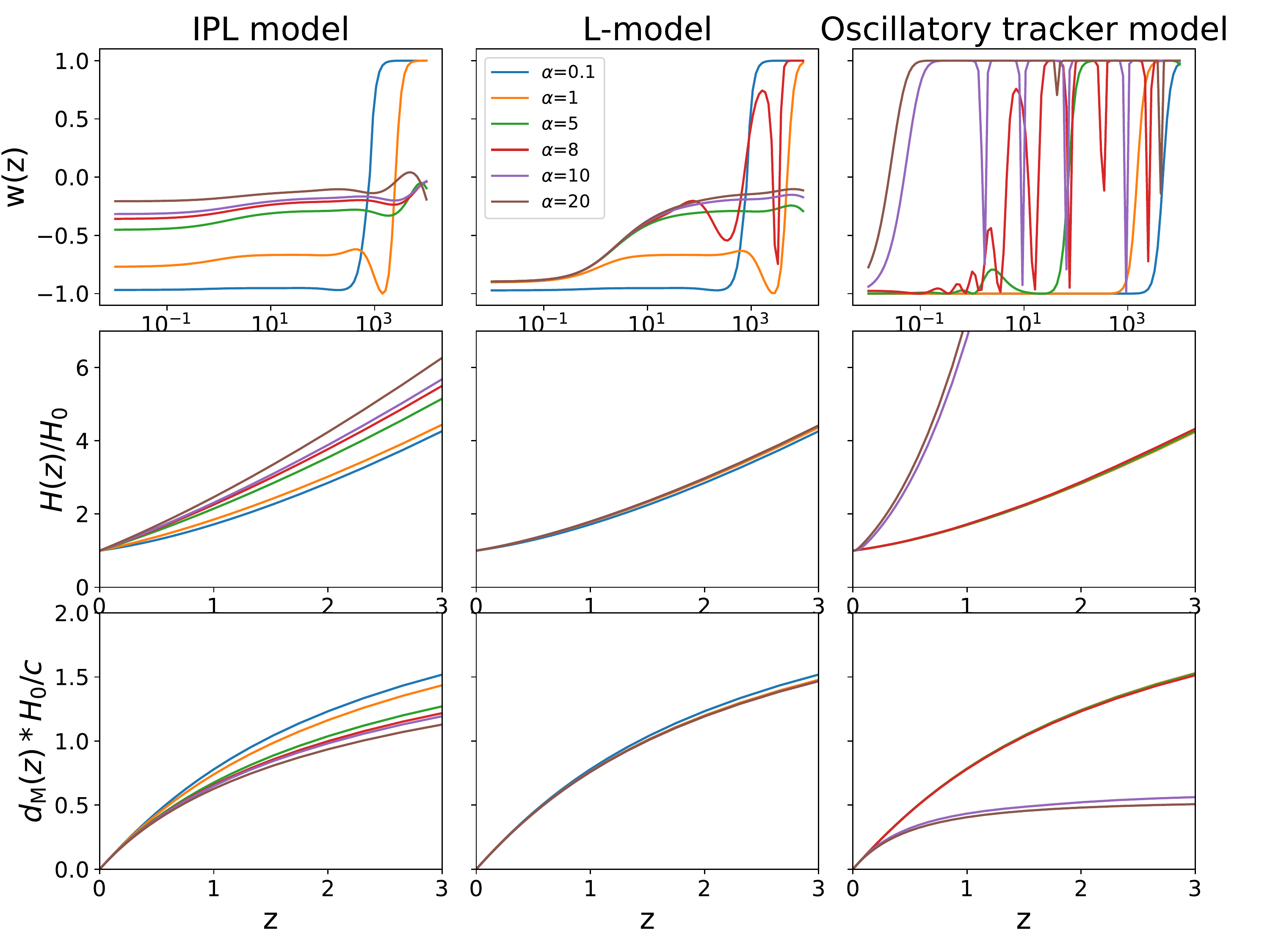}
		\caption{The evolution of the equation of state $w$, the dimensionless Hubble parameter $H/H_0$ and the transverse comoving distance $d_{\mathrm M}*H_0/c$ with respect to the redshift $z$ with the variation of parameter $\alpha$, while fixing $(\Omega_{\rm m0}, \Omega_{\rm k0})=(0.27, 0)$.}
		\label{fig:whdz}
	\end{figure}
	
	The SNe Ia data, BAO data, and $\emph{Planck}$ data, which have been introduced in Section 3, are used to constrain the models under consideration. The joint likelihood for parameters $\textbf{p}$ is:
	\begin{equation}
		\mathcal{L}(\textbf{p}) = \prod_i \mathcal{L}_i,
	\end{equation}
	where $\mathcal{L}_i$ is the likelihood of each data set used in the analysis. For example, the likelihood of the combining sample (SNe Ia+BAO+\emph{Planck}) for parameter set $\textbf{p}$ is:
	\begin{equation}
		\mathcal{L}(\textbf{p}) = \mathcal{L}_{\rm SNe}\mathcal{L}_{\rm BAO}\mathcal{L}_{\emph{Planck}}.
		\label{eq:likelihood}
	\end{equation}
	Then the effective $\chi^2$ is defined as:
	\begin{equation}
	    \chi^2_{\rm eff} = -2 * \ln{\mathcal{L}}.
	\end{equation}
	We use the same priors for the parameters as those from the $\emph{Planck}$ 2018 paper \cite{1807.06209} and adopt flat priors over reasonable ranges for the remaining parameters. An affine–invariant Markov chain Monte Carlo (MCMC) ensemble sampler (emcee)\cite{1202.3665} is employed to generate the posterior probability distributions for the parameters.

	\subsection{Observational constraints}
	The marginalized 2-D posterior distributions of the parameters ($\Omega_{\mathrm m 0}$, $\Omega_{\mathrm k 0}$, $\alpha$) for the three $\phi$CDM models are shown in Figure \ref{fig:results} and Figure \ref{fig:results_planck}, while the mean values with 68.3\% confidence limits for the parameters and the minimum $\chi^2_{\rm eff}$ of the combining sample are shown in Table \ref{tab:results}.
	The Figure \ref{fig:results} shows the constraint results of the three $\phi$CDM models arising from the SNe Ia data or BAO data individually, as well as from the combination of SNe Ia + BAO. We notice that $\alpha$ can hardly be constrained by the SNe Ia data and/or BAO data in the L-model. It seems that the SNe Ia and BAO data are both weak at constraining the parameter $\alpha$ in the L-model. To explore the reason behind it, we calculate the equation of state (EoS) $w(z)$ of the field $\phi$ within $0<z<10^4$, the dimensionless Hubble parameter $H(z)/H_0$ and the dimensionless transverse comoving distance $d_{\rm M}(z)*H_0/c$ within $0<z<3$ (which covers all redshifts of SNe Ia and BAO data) with taking different values of $\alpha$ in the three $\phi$CDM models, and the results are displayed in Figure \ref{fig:whdz}. 
	
	The first column of Figure \ref{fig:whdz}, corresponding to the IPL model, shows that the best-fit value $\alpha=0.22$ lies in the range where $w(z)$, $H(z)$ and $d_{\mathrm M}(z)$ are all sensitive to the variation of $\alpha$, which leads to the tight constraint result for $\alpha$. The second column of Figure \ref{fig:whdz}, corresponding to the L-model, illustrates that the evolution of the EoS are similar with the IPL model at early time, which is consistent with our inference in Section 2. One can notice that the difference between IPL model and L-model are mainly reflected in the $w(z)$ evolution at late time (approximately $z<100$).
	The EoS $w(z)$ at present ($z=0$) with different $\alpha$ are all closed to -1, which cause that $H(z)$ and $d_{\mathrm M}(z)$ are thoroughly insensitive to the variation of $\alpha$. Since the theoretical predictions for the observables of the SNe Ia data and BAO data depend solely on $H(z)$ and $d_{\mathrm M}(z)$ at $0<z<3$, the upper limit of the parameter $\alpha$ cannot be worked out in the L-model. The third column of Figure \ref{fig:whdz} corresponds to the Oscillatory tracker model. The oscillatory property of this model is shown obviously in the $w(z)$ panel. We notice that $w(z)$ converges to -1 at present epoch while $\alpha$ is approximately less than 8, that cause the similar $H(z)$ and $d_{\mathrm M}(z)$ curves at $0<z<3$, which explains the flatness of the parameter $\alpha$ within $0<\alpha<8$ in Fig \ref{fig:results}. While $\alpha$ is large than 8, $w(z)$ no longer converges to -1 at present epoch, such that the $H(z)$ and the $d_{\mathrm M}(z)$ curves present huge differences with $\alpha<8$. Therefore we conclude that, if the theoretical prediction for the observable of the data depends solely on $H(z)$ or $d_{\rm M}(z)$, the parameter $\alpha$ in the L-model cannot be constrained by the data, and the Oscillatory tracker model cannot distinguishes $\alpha$ at $\alpha<8$, while $\alpha$ in the IPL model can be strictly constrained.
	
	Fortunately, the power spectra of the CMB temperature and polarization anisotropies from the $\emph{Planck}$ data contain more cosmological information than the other common cosmological probes. The results in Figure \ref{fig:results_planck} show that the $\emph{Planck}$ data place remarkable constraints on all parameters (including $\alpha$) of the three $\phi$CDM models. When combining the $\emph{Planck}$ data with SNe Ia and BAO data, the constraint results are more strict, and the degeneracy among model parameters are well broken. The results from the combining sample denote that all these three $\phi$CDM models do not exclude the $\Lambda$CDM model at $68.3\%$ confidence interval.
	
	\subsection{Bayesian model selection}
	
	According to the Bayes' theorem \cite{0803.4089}, under the condition of observational data $d$, the probability that cosmological model $\mathcal{M}$ is true can be written as:
	
	\begin{equation}
		\begin{aligned}
			P(\mathcal{M}|d) &=\frac{P(d|\mathcal{M})}{P(d)}P(\mathcal{M}) \\
			&= \frac{P(d|\mathcal{M})}{\Sigma P(d|\mathcal{M}_i)P(\mathcal{M}_i)}P(\mathcal{M}),
		\end{aligned}
	\end{equation}
	where $P(\mathcal{M})$ and $P(\mathcal{M}|d)$ are the prior probability and posterior probability, respectively.
	
	If there are two cosmological models $M_0$ and $M_1$, to determine which model is much preferred by the observational data, one can calculate the ratio of posterior probability (or posterior odds) of the two models:
	\begin{equation}
		\begin{aligned}
			\frac{P(\mathcal{M}_1|d)}{P(\mathcal{M}_0|d)}&=\frac{P(d|\mathcal{M}_1)P(\mathcal{M}_1)}{P(d|\mathcal{M}_0)P(\mathcal{M}_0)} \\
			&=B_{10}\frac{P(\mathcal{M}_1)}{P(\mathcal{M}_0)},
		\end{aligned}
		\label{eq:probability}
	\end{equation}
	where $B_{10}$ is the Bayes factor between $\mathcal{M}_1$ and $\mathcal{M}_0$:
	\begin{equation}
		B_{10} \equiv \frac{P(d|\mathcal{M}_1)}{P(d|\mathcal{M}_0)},
	\end{equation}
	which is defined as the ratio of the Bayesian evidences $P(d|\mathcal{M})$ of the two models. The Eq(\ref{eq:probability}) indicates that	the posteriors odds equals the prior odds times the Bayes factor, where the Bayes factor $B_{10}>1$ or $<1$ reflects whether the observational data prefer $\mathcal{M}_1$ rather than $\mathcal{M}_0$ or not. In addition, the ``Jeffreys' scale'' \cite{0803.4089} suggests that, $\lvert \ln{B_{10}} \rvert \in (0, 1.0), (1.0, 2.5), (2.5, 5.0), (5.0, \infty)$ correspond to the inclusive, weak, moderate, and strong evidences, respectively. 
	
\begin{figure}
    \centering
    \subfigure[IPL model]{
    \begin{minipage}[b]{0.45\textwidth}
    \includegraphics[width=\textwidth]{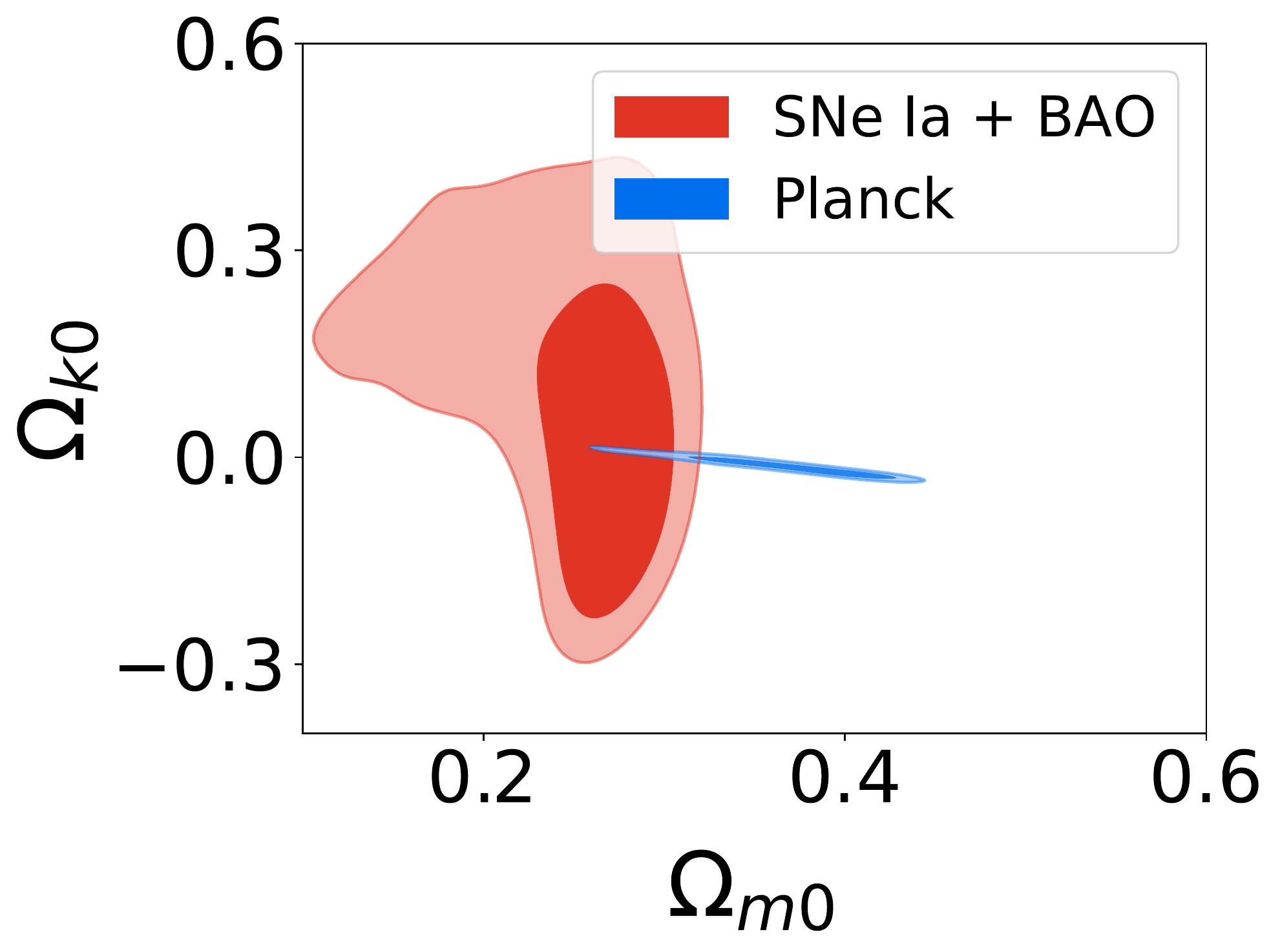}
    \end{minipage}
    }
    \subfigure[L-model]{
    \begin{minipage}[b]{0.45\textwidth}
    \includegraphics[width=\textwidth]{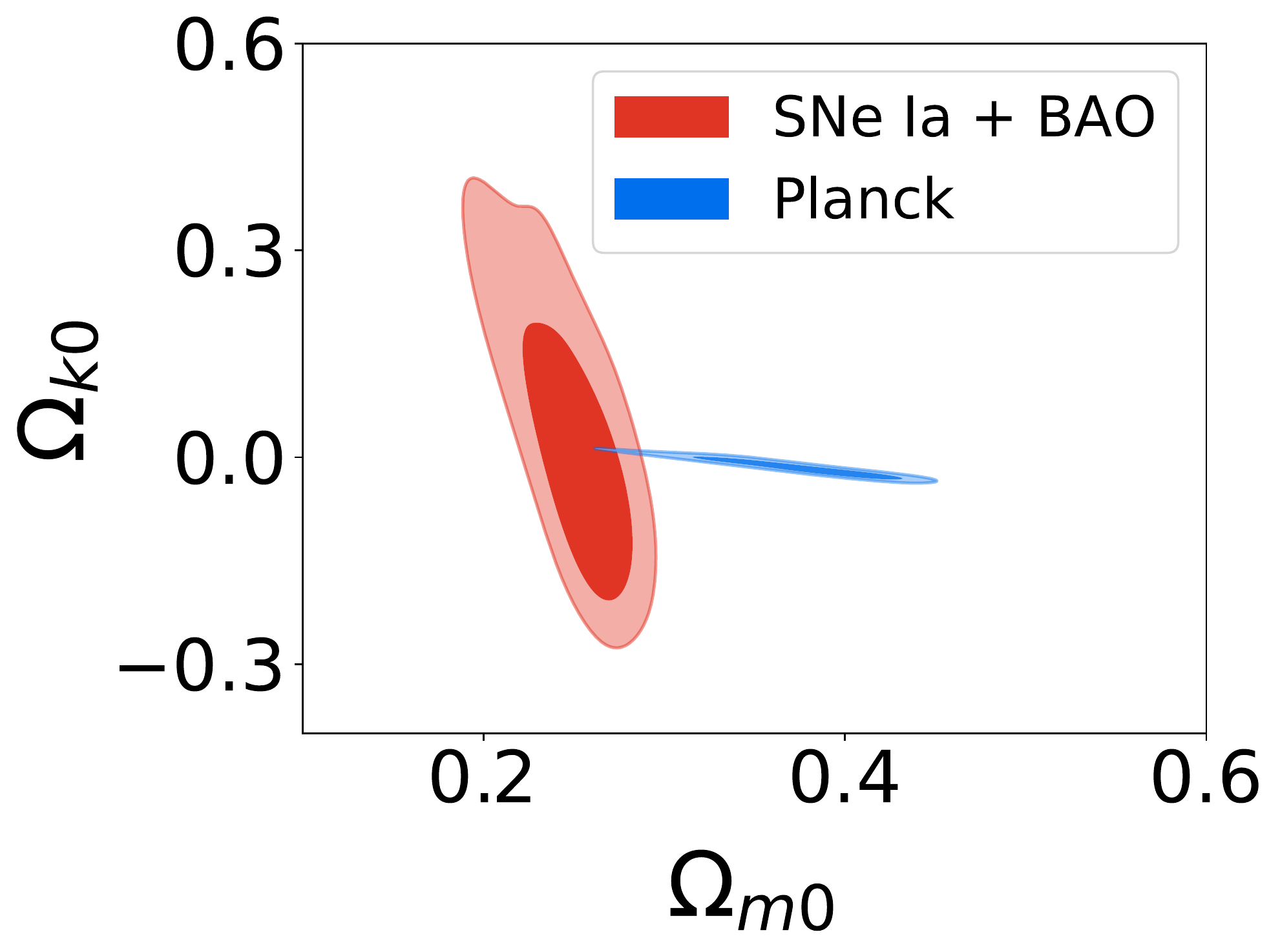}
    \end{minipage}
    }
    \caption{
   Contours refer to the marginalized likelihoods at $68.3\%$ and $95.4\%$ confidence levels in the  ($\Omega_{\rm m0}, \Omega_{\mathrm{k0}}$) plane for the IPL model and  L-model respectively, constrained from the SNe Ia+BAO data and the \emph{Planck} data .}
    
    \label{fig:OmOk}
    \end{figure}

	In section 4.1, we have used the SNe Ia, BAO and \emph{Planck} data to constrain the three $\phi$CDM models. Here we apply Bayes’ theorem to compare the
	$\phi$CDM models with the $\Lambda$CDM model. For this purpose, we should calculate the Bayesian evidence $P(d|\mathcal{M})$ of each model $\mathcal{M}$. The code \href{https://github.com/yabebalFantaye/MCEvidence/tree/master}{\textbf{MCEvidence}}\footnote{https://github.com/yabebalFantaye/MCEvidence/tree/master} \cite{1704.03472}, which is widely used to calculate the Bayesian evidence, is employed here.
	In Table \ref{tab:results}, we show the natural logarithm of the Bayesian evidence, which is denoted as $B_i$, of the three $\phi$CDM models and the $\Lambda$CDM model, as well as the Bayes factor $B_{i0}$, where $i=1,2$ and 3 are used to denote the three $\phi$CDM models and 0 represents the $\Lambda$CDM model. It turns out that the $\Lambda$CDM model is most favoured by the combining sample compared with the three $\phi$CDM models, since the Bayes factors of the three $\phi$CDM are all less than 1. Furthermore, $\lvert \ln{B_{10}} \rvert =0.66$ is in the range $(0, 1.0)$, which indicates that there is an inconclusive evidence that the observational data prefer the $\Lambda$CDM model rather than the IPL model according to the ``Jeffreys' scale'' \cite{0803.4089} mentioned above. Similarly, $ \lvert \ln{B_{20}} \rvert \sim 5.0$ infers that the observational data prefer the $\Lambda$CDM model rather than the L-model,	while $1.0< \lvert \ln{B_{30}} \rvert <2.5$ implies that there is a weak evidence that the observational data prefer the $\Lambda$CDM model rather than the Oscillatory tracker model. From Table \ref{tab:results}, we can see that the difference of the Bayes factors between the IPL model and the L-model mainly comes from the difference of the $\chi^2_{\mathrm{min}}$. We have investigated the reason of why the $\chi^2_{\mathrm{min}}$ of the L-model constrained from the combining sample is significantly larger than that of the IPL model. We find that the value of $\chi^2_{\mathrm{min}}$ constrained from the SNe Ia+BAO data or the \emph{Planck} data in the L-model is very close to that in the IPL model, but the overlapping degree of the 95.4\% confidence intervals of $\Omega_{\mathrm m0}$ constrained from the SNe Ia+BAO data and the \emph{Planck} data is much lower in the L-model than that in the IPL model (see Figure \ref{fig:OmOk}), hence the combining sample produces a larger $\chi^2_{\mathrm{min}}$ in the L-model.
	
	Linares Cede{\~n}o et al.(2019) \cite{1905.00834} have conducted observational constraints on several $\phi$CDM models including an another form of L-model, i.e., $V(\phi)\propto \coth{(\alpha\phi)}$, and the Oscillatory tracker model same with this work, with the observational Hubble parameter data (OHD) derived from the differential evolution of massive and passive early-type galaxies, the SNe Ia data from Pantheon compilation, the BAO data from SDSS DR12, 6dFGS and SDSS DR7, and the CMB data in the condensed form of shift parameters derived from the \emph{Planck} 2018 results, and also has used the code
	\href{https://github.com/yabebalFantaye/MCEvidence/tree/master}{\textbf{MCEvidence}} \cite{1704.03472} to estimate the Bayesian evidence for these models. The confidence intervals of model parameters are not displayed in their work, but they have presented the Bayesian evidence for each model. Their analyses based on the Bayesian evidence state that the Oscillatory Tracker model is weakly favored compared with the $\Lambda$CDM model, which is different from our results. We speculate that the main reason for the difference is that the CMB data they have used are the distance priors rather than the CMB power spectra data used in our analysis.
	
	\section{Conclusions and discussions}
	
	In this paper, we study and compare three kinds of scalar field dark energy models ($\phi$CDM models) with tracker properties by using the recent SNe Ia, BAO and CMB data, which own the potentials $V(\phi)\propto \phi^{-\alpha}$ (IPL model), $V(\phi)\propto \coth^{\alpha}{\phi}$ (L-model) and $V(\phi)\propto \cosh(\alpha\phi)$ (Oscillatory tracker model), respectively. For the L-model, we adopt a subclass of the L-model introduced in Satadru et al. (2017) \cite{1709.09193}. There are two parameters we are interested in, i.e., the current curvature density parameter $\Omega_{\mathrm{k} 0}$, and the parameter $\alpha$ which quantifies the discrepancy between each $\phi$CDM model and the $\Lambda$CDM model. We use the MCMC method to obtain the posterior probability distributions of the model parameters with the analysis of SNe Ia Pantheon data, BAO data from 6dFGS and SDSS, as well as the data of CMB temperature and polarization anisotropies from the \emph{Planck} 2018 legacy data release. Additionally, we apply the Bayesian evidence to compare the three $\phi$CDM models with the $\Lambda$CDM model. The main results are summarized as follows:
	
	\begin{itemize}
		
		\item The constraint results from the combining sample infer that none of the three $\phi$CDM models exclude the $\Lambda$CDM model at $68.3\%$ confidence level. 
		
		\item If the theoretical prediction for the observable of the data depends solely on $H(z)$ or $d_{\rm M}(z)$, e.g, SNe Ia data and BAO data, they are weak at constraining the parameter $\alpha$ in the L-model and the parameter $\alpha$ in the oscillatory tracker model within $0<\alpha<8$, while this problem dose not exist in the IPL model.
		
		\item The \emph{Planck} TT, EE and TE power spectra data can place strict constraints on the model parameters of the three $\phi$CDM models. Even for the parameter $\alpha$ in the L-model, which is weakly constrained by SNe Ia data or BAO data, the \emph{Planck} CMB power spectra data have an excellent performance. While further combined it with SNe Ia data and BAO data, the degeneracies among the model parameters are well broken and the constraints are more strict.
		
		\item The analysis results of the Bayesian evidence turn out that the $\Lambda$CDM model is the most competitive model compared with the three $\phi$CDM models, but the evidence that the $\Lambda$CDM model is better than the IPL model is inconclusive. Moreover, there is a weak evidence that the observational data prefer the $\Lambda$CDM model rather than the Oscillatory tracker model, while the evidence that L-model is disfavored by the observational data is strong.
	\end{itemize}
	
	The recent observations of standard cosmological probes are adopted to investigate the $\phi$CDM models under consideration. It would be interesting to further explore the powers of other cosmological probes on constraining and comparing these models.
	
	Even though our results show that the $\Lambda$CDM model is still the most preferred cosmological model by the observational data, we still get some useful information from this study. 
	On the one hand, The primary advantage of the cosmological models with quintessence is that they can give rise to $w_{\rm DE} \sim -1$ today from a large basin of attraction motivated by $\alpha$-attractors. They somewhat address the fine tuning problem while satisfying the current observations, e.g. SNe Ia, BAO and CMB. 
	On the other hand, it is known that the differences among various dark energy models are embodied in the evolution of the dark energy EoS ($w(z)$), and then further reflected in the Hubble-redshift relation ($H(z)$) and distance-redshift relation ($d(z)$). However, from the study of the L-model and the oscillatory tracker model, we find that though $w(z)$ vary with different values of the parameter $\alpha$, the corresponding $H(z)$ and $d(z)$ are still indistinguishable. This result indicates that the information contained in the cosmological probes solely based on $H(z)$ and/or $d(z)$ may not be complete, and they are not able to place sufficient observational constraints for some special dark energy models. Compared with these cosmological probes, the CMB temperature and polarization anisotropy power spectra data render more comprehensive cosmological information including the background evolution and the perturbation evolution in the early universe. The differences among the dark energy models are also reflected in the power spectra - even at the era that the dark energy is not the dominant component in the universe.

	From the perspective of model comparison, the simple BIC (Bayesian Information Criterion) \cite{astro-ph/0701113}  method  only make a rough judgement via including the contributions from the maximum likelihood value, the number of model parameters and the number of data points, while the Bayesian evidence applied in this work can compare the fitting results of the models under consideration more accurately and fairly. We further anticipate to explore more advanced methods to evaluate the fitting results of various cosmological models.
	
	\section*{Acknowledgments}
	
	This work has been supported by the National Natural Science Foundation of China (Nos. 11988101 and 12033008), and the K. C. Wong Education Foundation. LX is partially supported by the National Natural Science Foundation of China under the Grant Nos. 11675032 and 12075042. S. Cao is supported by the National Natural Science Foundation of China under Grant Nos. 12021003, 11690023, 11633001 and 11920101003, the National Key R\&D Program of China (Grant No. 2017YFA0402600), the Beijing Talents Fund of Organization Department of Beijing Municipal Committee of the CPC, the Strategic Priority Research Program of the Chinese Academy of Sciences (Grant No. XDB23000000), and the Interdiscipline Research Funds of Beijing Normal University.


\end{document}